\documentclass[a4paper,dvips,aps,preprint,
	superscriptaddress, tightenlines, showpacs,byrevtex]{revtex4}

\usepackage{graphicx}
\graphicspath{{ps/}}
\usepackage{psfrag}
\usepackage[usenames]{color}

\begin{document}

\title{\Large
	\vspace*{2ex}
	Measurement of $D^0$ decays to $\KL \pi^0$ and $\KS \pi^0$ at Belle
	\\
	\vspace*{2ex}
      }

\preprint{BELLE-CONF-0129}


\newcommand{\GeV}{\ensuremath{\mathrm{GeV}}}
\newcommand{\MeV}{\ensuremath{\mathrm{MeV}}}

\newcommand{\KL}{\ensuremath{K^0_L}}
\newcommand{\pKL}{\ensuremath{\widetilde{K^0_L}}}
\newcommand{\KS}{\ensuremath{K^0_S}}
\newcommand{\Kstar}{\ensuremath{K^{\ast -}}}
\newcommand{\Dstar}{\ensuremath{D^{\ast +}}}
\newcommand{\thDK}{\ensuremath{\theta_{DK}}}

\begin{abstract}
We present a preliminary measurement of the ratio of $D^0$ decay rates
into $\KL \pi^0$ and $\KS \pi^0$ final states. This ratio can be used
to disentangle the Cabibbo favored $D^0 \rightarrow \overline{K}^0
\pi^0$ and doubly Cabibbo suppressed $D^0 \rightarrow K^0 \pi^0$
amplitudes, and contributes to the important goal of constraining the
strong phase $\delta_{K \pi}$ between $D^0 \rightarrow K^- \pi^+$ and
$D^0 \rightarrow K^+ \pi^-$. The measurement is based on a large data
set accumulated by the Belle detector at the KEKB $e^+ e^-$ collider,
with $\KL$ candidates reconstructed using hadronic clusters in the
KLM (\KL\ and $\mu$ identification system), together with a $D^0$ mass
constraint and $D^\ast$ tag.
\end{abstract}

\author{The Belle Collaboration}

\noaffiliation

\maketitle


\begin{center}
  K.~Abe$^{9}$,               
  K.~Abe$^{37}$,              
  R.~Abe$^{27}$,              
  I.~Adachi$^{9}$,            
  Byoung~Sup~Ahn$^{16}$,      
  H.~Aihara$^{39}$,           
  M.~Akatsu$^{20}$,           
  K.~Asai$^{21}$,             
  M.~Asai$^{10}$,             
  Y.~Asano$^{44}$
% ...
\end{center}

\small
\begin{center}
% ...
$^{9}${High Energy Accelerator Research Organization (KEK), Tsukuba}\\
$^{10}${Hiroshima Institute of Technology, Hiroshima}\\
% ...
$^{16}${Korea University, Seoul}\\
% ...
$^{20}${Nagoya University, Nagoya}\\
$^{21}${Nara Women's University, Nara}\\
% ...
$^{27}${Niigata University, Niigata}\\
% ...
$^{37}${Tohoku Gakuin University, Tagajo}\\
% ...
$^{39}${University of Tokyo, Tokyo}\\
% ...
$^{44}${University of Tsukuba, Tsukuba}\\
% ...
\end{center}


\section{Introduction}

The search for $D^0 - \overline{D}^0$ mixing has long been one of the
most interesting projects in charm physics. Recent publications by the
FOCUS~\cite{FOCUS} and CLEO~\cite{CLEO} collaborations have stimulated
a series of new measurements, including an updated $y_{CP}$
measurement by the Belle collaboration~\cite{Belle:y_CP}, reported at
the summer conferences. However, the experimental situation is still
not clear.

An important quantity in the interpretation of $D^0 - \overline{D}^0$
mixing results is the phase difference $\delta_{K \pi}$ between the
Cabibbo favored (CF) and doubly Cabibbo suppressed (DCS) neutral $D$
meson decays.  As has been shown in~\cite{Golowich}, new information
on $\delta_{K \pi}$ may be obtained by measuring the asymmetry between
the decay rates of $D^0$ into $\KS \pi^0$ and $\KL \pi^0$, where the
effect may be as large as $\mathcal{O}(\tan^2 \theta_c)$, i.e. at the 5\%
level.

In this paper we present the first measurement of the $D^0 \rightarrow
\KL \pi^0 / D^0 \rightarrow \KS \pi^0$ decay rate asymmetry, using the
Belle detector at the KEKB $e^+ e^-$ collider. In order to cancel out
most of the systematic effects of the detector, we extract the $\KL$/$\KS$
relative detection efficiency from the ratio of $\KL \pi^+ \pi^-$ and $\KS
\pi^+ \pi^-$ modes via the decay $D^0 \rightarrow \Kstar \pi^+$.
The presence of $\Kstar \rightarrow K^0 \pi^-$ in the decay chain ensures equal
rates of $\KL$ and $\KS$ in this case.

Throughout this paper, the inclusion of CP-conjugate processes is implied.
In particular, we combine $D^0 \rightarrow \KL \pi^0$ with
$\overline{D}^0 \rightarrow \KL \pi^0$, and so on.
We thus treat the \KL\ and \KS\ as CP-eigenstates (the error is negligible)
and assume that direct CP violation in $D \rightarrow \overline{K}\pi$ and
$K \pi$ decays, if present, may be ignored at our present sensitivity.


\section{$\KL$ reconstruction technique}

The key point in the analysis of final states containing $\KL$ is the
reconstruction of its momentum.  The technique presented below makes
use of the KLM (\KL\ and $\mu$ identification) subsystem of the Belle
detector~\cite{Belle:detector}, which consists of several layers of
iron absorber interspaced with resistive plate chambers.  The system
enables one to reconstruct muon tracks as well as hadronic showers.
Those showers which do not have a matching charged track are assumed
to be produced by $\KL$, with the position of the shower giving
information about the flight direction of the $\KL$.  The direction
resolution is improved if the $\KL$ deposits energy in the
electromagnetic calorimeter in front of the KLM, in which case the
position of the calorimeter cluster is used to determine the $\KL$
flight direction.

To extract the \emph{magnitude} of the $\KL$ momentum one needs to
employ additional constraints.  In particular to reconstruct $D^0$
decays into $\KL X$ where $X$ is a fully reconstructed system, one can
apply a $D^0$ mass constraint and solve the resulting 4-momentum
equation with respect to the magnitude of the $\KL$ momentum, and then exploit
the $D^{\ast} \rightarrow D^0 \pi$ decay to tag the signal.

To monitor the performance of the method we utilize the same $D^0$
decay modes but with $\KS$ instead of $\KL$ in the final state.  We
use the $\KS$ flight direction reconstructed by its decay into $\pi^+
\pi^-$, and then apply the same procedure as for $\KL$: henceforth
such $\KS$ are referred to as $\pKL$ (pseudo $\KL$).

Since the equation for the magnitude of the $\KL$ momentum is quadratic
there may be up to two solutions.  However one of those, even if present,
is non--physical in more than $90\%$ of events. For simplicity,
in this analysis we use only the one with the larger momentum.


\section{Data set and event selection}

The data sample used for this analysis amounts in total to an
integrated luminosity of $23.6 \mathrm{~fb}^{-1}$.

We select $D^\ast$ candidates with reconstructed scaled momentum $x_p > 0.6$,
which rejects $D^\ast$ from $B$-meson decays
and supresses the combinatoric background.
In addition to the standard reconstruction procedure for $\pi^\pm$,
$\pi^0$, and $\KS$, the latter are required to satisfy the
following quality criteria:
\begin{itemize}
\item the $\KS \rightarrow \pi^+ \pi^-$ vertex should be separated from the interaction point
in the plane perpendicular to the beam axis by more than $500~\mu\mathrm{m}$;
\item the distance along the beam axis between the $\pi^\pm$ tracks at the
\KS\ vertex must be less than $1 \mathrm{~cm}$;
\item the cosine of the angle between the assumed $\KS$ flight path
from the interaction point to the decay vertex and the reconstructed
$\KS$ momentum in the transverse projection with respect to the beam must be
more than 0.95.
\end{itemize}
When reconstructing \KL, there is a potential background due to KLM clusters
caused by unreconstructed charged particles, in particular when these
particles do not originate from the interaction region
(e.g. $K$ and $\pi$ decays in flight). To reject such cases we veto any KLM
cluster matched to a calorimeter cluster with an energy between
$0.15$ and $0.3~\GeV$, corresponding to minimum ionization.

\begin{figure}[htbp]
\begin{center}
\psfrag{xLabel}[tr][tr]{$\cos{\thDK}$}
\psfrag{yLabel}[cr][cr]{$\left[\frac{1}{0.05}\right]$}
\begin{tabular}{@{}c@{}c@{}}
\includegraphics[height=.2\textheight,width=.48\textwidth]{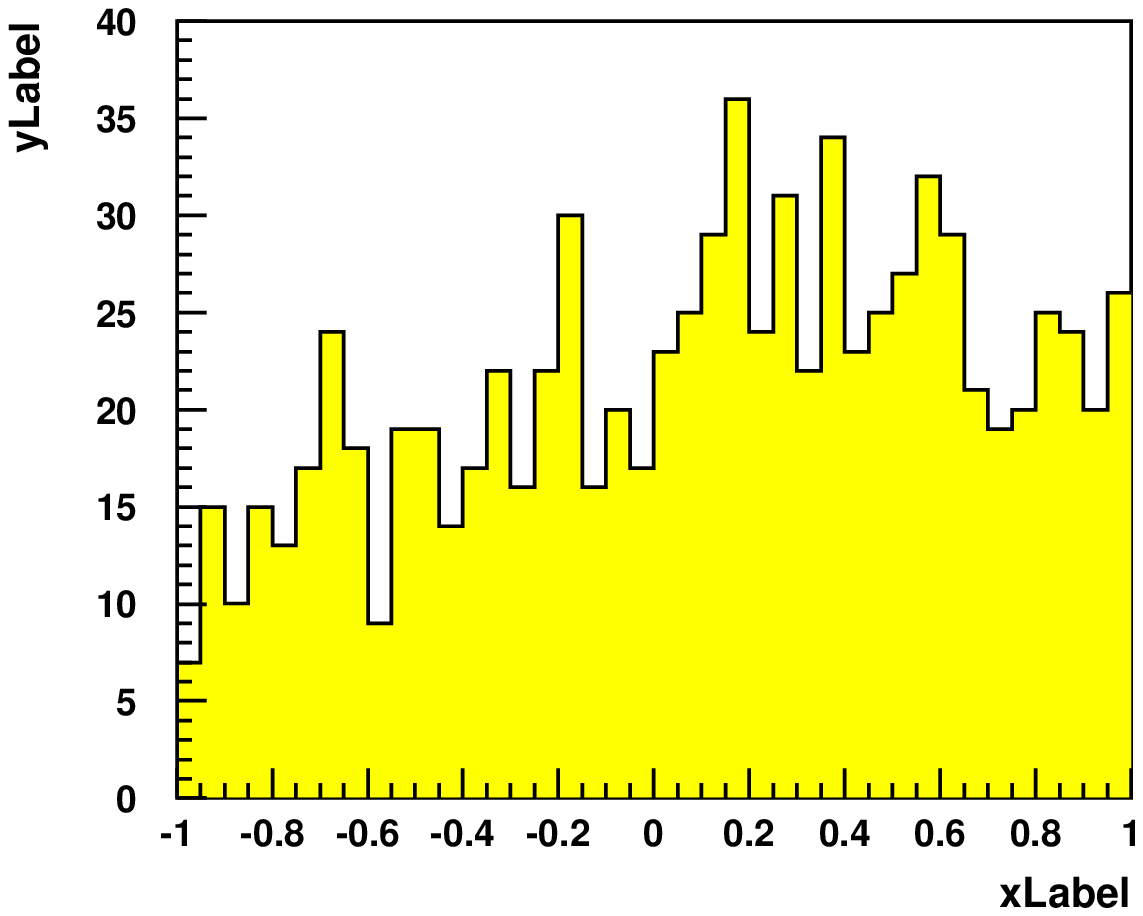} &
\includegraphics[height=.2\textheight,width=.48\textwidth]{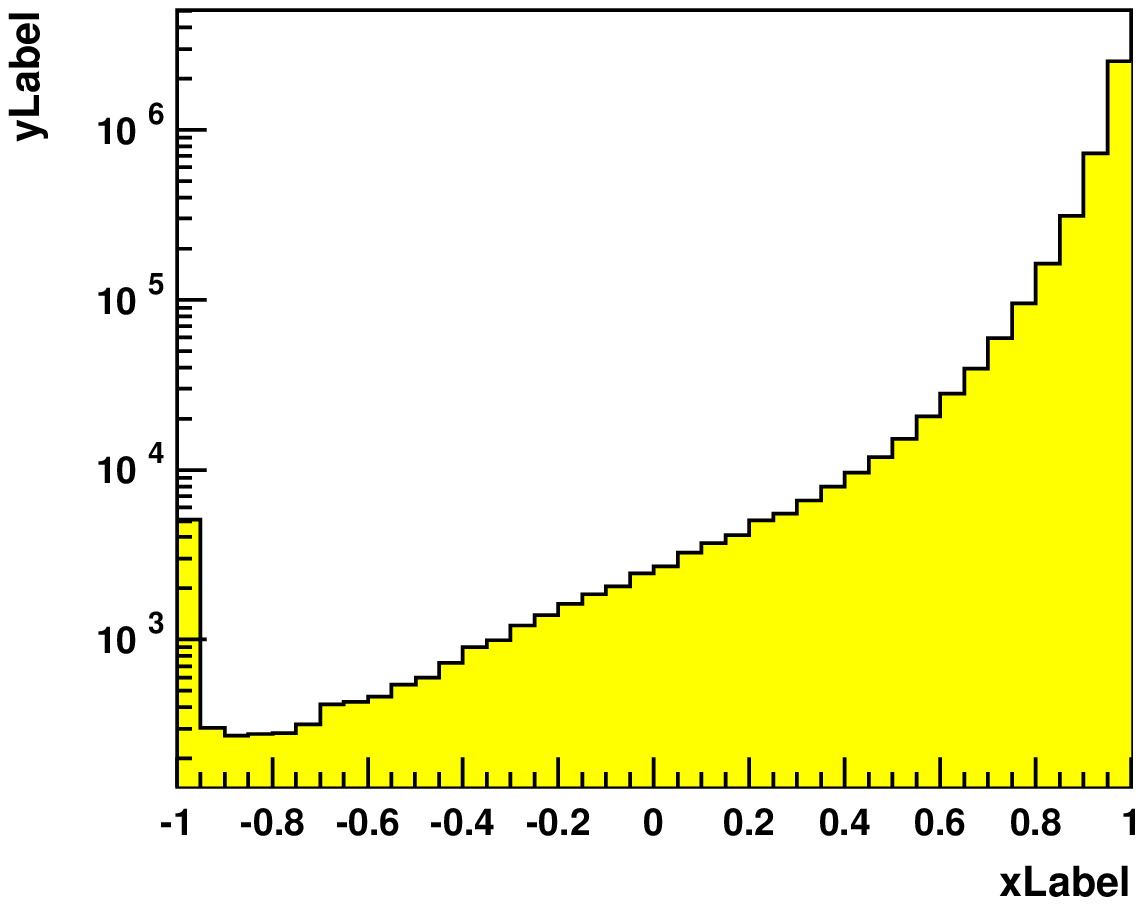} \\
a) signal MC &
b) $M(D^\ast)$ sideband data
\end{tabular}
\end{center}
\caption{$K^0$ flight angle with respect to the $D^0$ boost for $D^0
\rightarrow \pKL \pi^0$}
\label{fig:cosD0K0}
\end{figure}

The dominant part of the background is due to the pickup of arbitrary
soft pions when forming a $D^0$ candidate.  A characteristic feature
of such combinations is that the $D^0$ mass is accommodated by
assigning a large momentum to the $K^0$, which then accounts for most
of the $D^0$ momentum.  Therefore the $K^0$ flight angle with respect to the
$D^0$ boost (hereafter $\thDK$) for this background tends
to peak in the forward direction (Fig.~\ref{fig:cosD0K0}b) while for
the signal the underlying distribution is isotropic, with a slight
tilt to the forward direction in the measured distribution
(Fig.~\ref{fig:cosD0K0}a) due to detector efficiency.

The sharp peak around $\cos{\thDK} = -1$ is due to the degenerate case
when a random $\pi^0$ or $\pi^+ \pi^-$ system can form
a ``good'' $D^0$ with a $K^0$ (almost) at rest in the lab: to
exclude these cases, we require $\cos{\thDK} > -0.95$. The upper
cut on this quantity needs optimization which has been performed using
signal MC and data in the $M(D^\ast)$ sideband: we require
$\cos{\thDK} < 0.2$ for all four modes.

The invariant mass of the $K^0 \pi^-$ combination in the $K^0 \pi^+ \pi^-$ mode
is required to lie
within $50~\MeV/c^2$ of the nominal $K^\ast(892)$ mass.  In
addition the invariant mass of the $\pi^+ \pi^-$ pair is required to
be less than $0.7~\GeV/c^2$ to make the signal and the control modes
kinematically similar, and to suppress the contribution from $K^0
\rho$ decays.

\begin{figure}[htbp]
\begin{center}
\psfrag{xLabel}[tr][tr]{$p^{lab}_{\KS} \left[\GeV/c\right]$}
\psfrag{yLabel}[br][br]{$\left[\frac{1}{0.1~\GeV/c}\right]$}
\begin{tabular}{c}
\includegraphics[height=.2\textheight,width=.48\textwidth]
{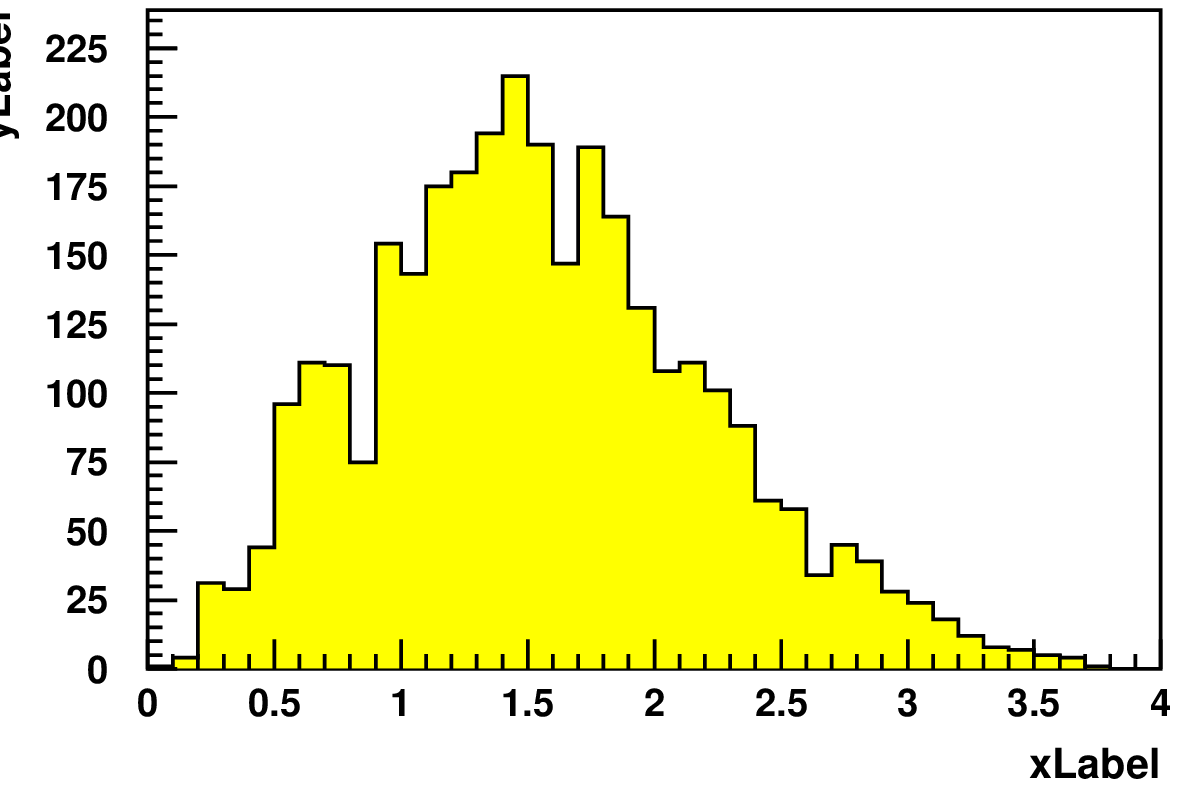}\\
(a) $\KS \pi^0$\\
\includegraphics[height=.2\textheight,width=.48\textwidth]
{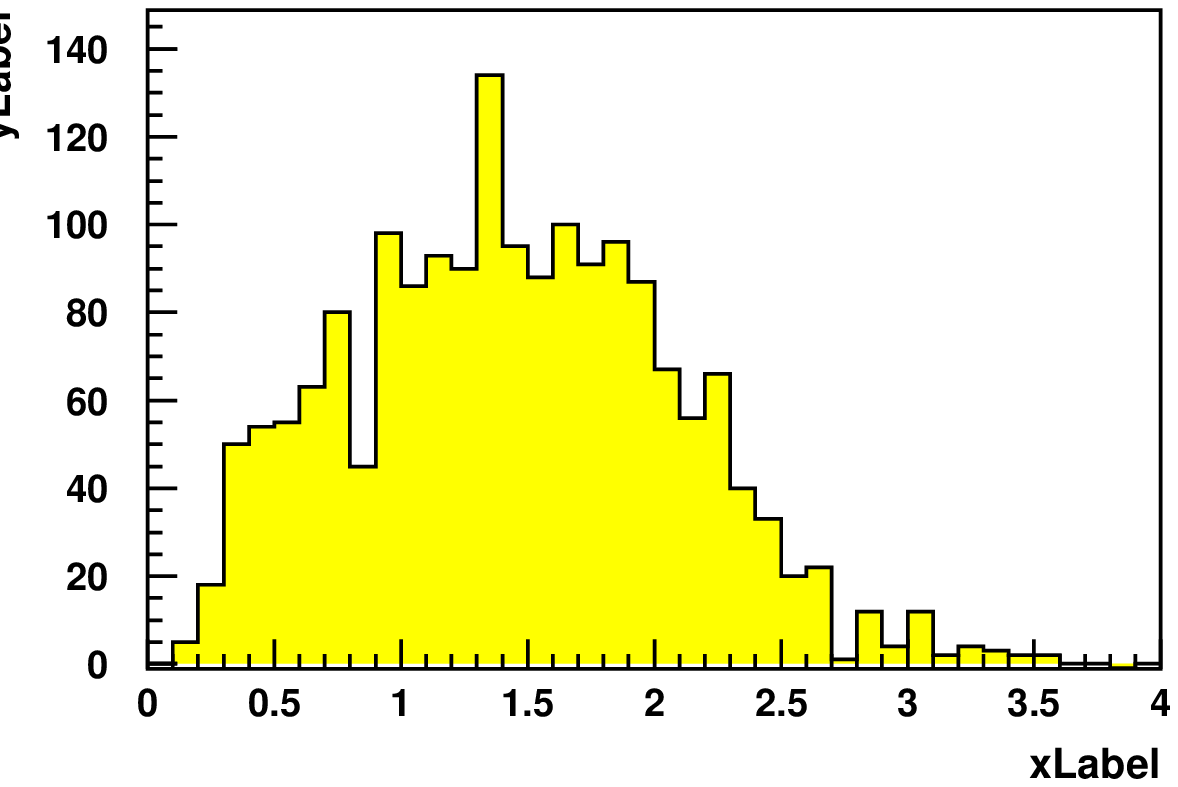}\\
(b) $(\KS \pi) \pi$\\
\psfrag{yLabel}[br][br]{$\left[\frac{1}{0.1~\GeV/c}\right]$}
\includegraphics[height=.2\textheight,width=.48\textwidth]
{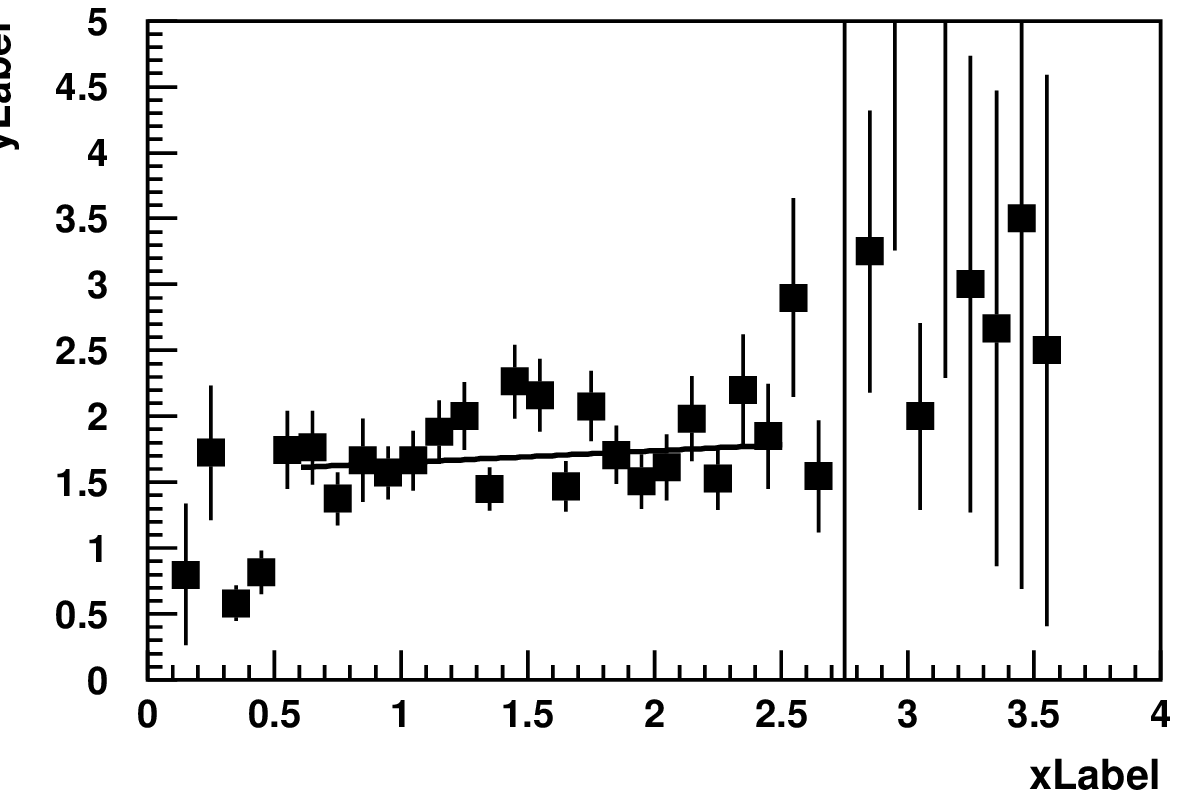} \\
(c) = (a) / (b)
\end{tabular}
\end{center}
\caption{$K^0$ lab momentum spectra for (a) $\KS \pi^0$ and (b) $(\KS \pi)\pi$;
	(c) shows the ratio of (a) to (b) as a function of momentum.}
\label{fig:PLspectra}
\end{figure}

Since the kinematics in the $\KL$ and the corresponding $\KS$ modes is
the same we can use the fully controlled $\KS \pi^0$ and $\KS \pi \pi$
modes to select a region in the $K^0$ lab momentum where the
efficiency of the kinematical cuts, which is independent of the $K^0$
flavor, is the same (see Fig.~\ref{fig:PLspectra}).  Hence we can
average over this region and all the efficiencies including that of
$\KL$ reconstruction cancel out.
We select the $K^0$ lab momentum range from
$0.6~\GeV/c$ to $2.5~\GeV/c$ for this analysis.

\section{Determination of $\KL$ and $\KS$ rate ratio}

\begin{figure}[htbp]
\begin{center}
\psfrag{Label}{}
\psfrag{yLabel}[br][br]{$\left[\frac{1}{0.25~\MeV/c^2}\right]$}
\begin{tabular}{@{}c@{\hspace{.04\textwidth}}c@{}}
\psfrag{xLabel}[tr][tr]{$M(D^0 \pi) - M(D^0) + M_{D^0}^{PDG} [\GeV/c^2]$}
\psfrag{nEvents}[bl][bl]{\fbox{$4715 \pm 91$}}
\includegraphics[height=.2\textheight,width=.46\textwidth]{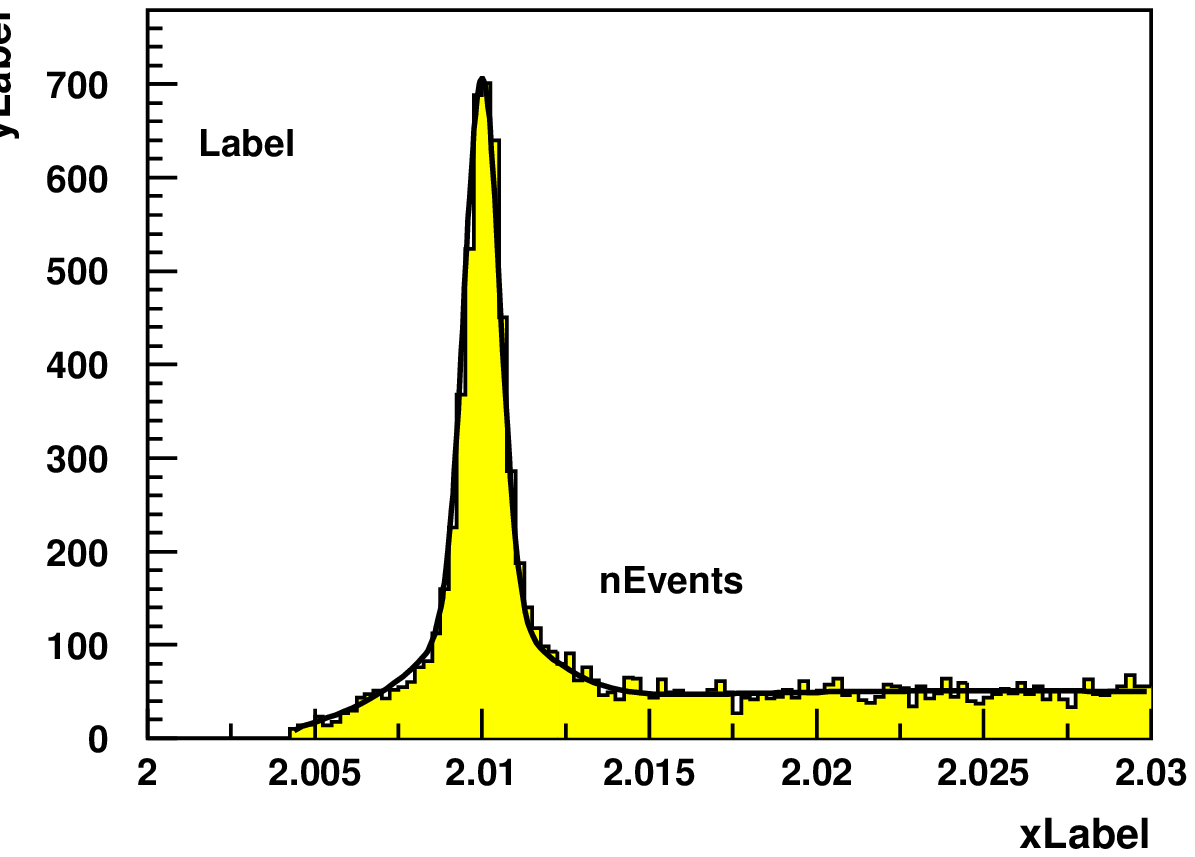} &
\psfrag{xLabel}[tr][tr]{$M(D^0 \pi) [\GeV/c^2]$}
\psfrag{nEvents}[bl][bl]{\fbox{$1839 \pm 101$}}
\includegraphics[height=.2\textheight,width=.46\textwidth]{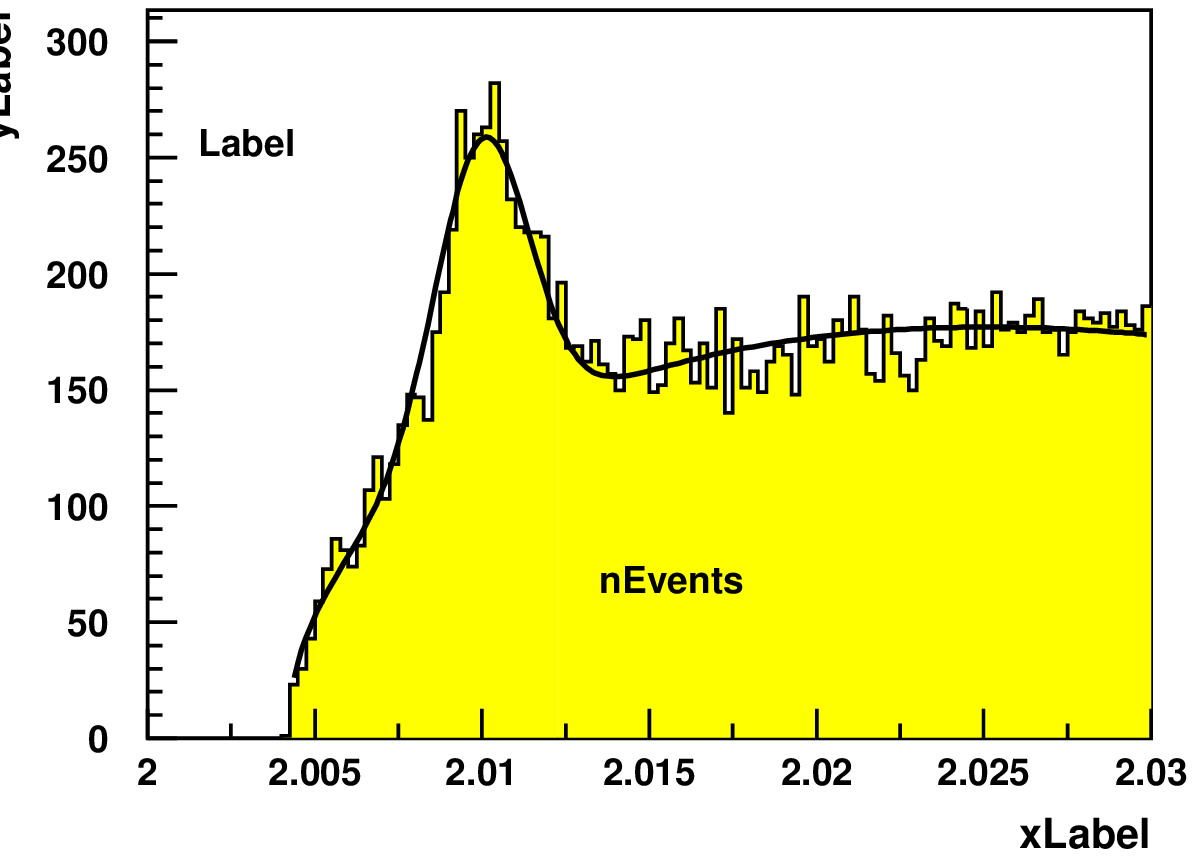} \\
a) $\KS \pi^0$ &
b) $\KL \pi^0$ \\\\
\psfrag{xLabel}[tr][tr]{$M(D^0 \pi) - M(D^0) + M_{D^0}^{PDG} [\GeV/c^2]$}
\psfrag{nEvents}[bl][bl]{\fbox{$2524 \pm 77$}}
\includegraphics[height=.2\textheight,width=.46\textwidth]{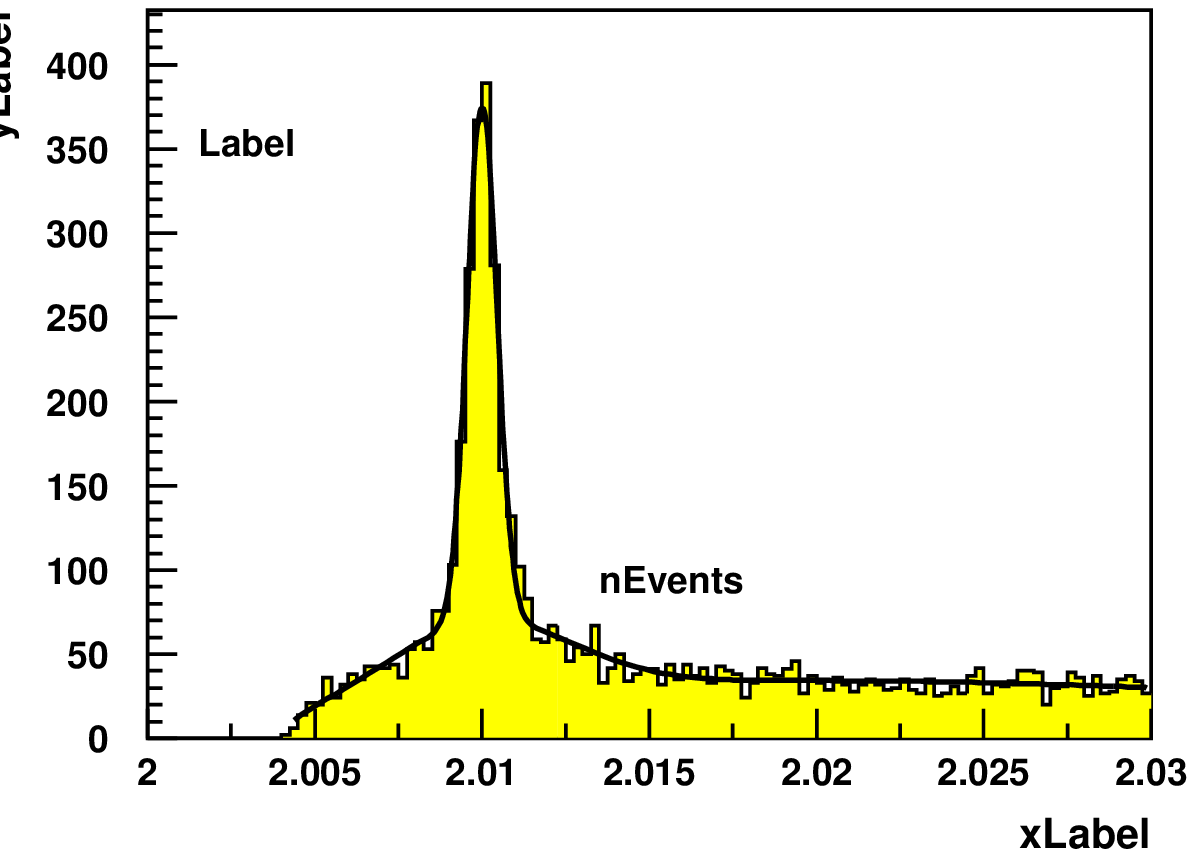} &
\psfrag{xLabel}[tr][tr]{$M(D^0 \pi) [\GeV/c^2]$}
\psfrag{nEvents}[bl][bl]{\fbox{$1119 \pm 84$}}
\includegraphics[height=.2\textheight,width=.46\textwidth]{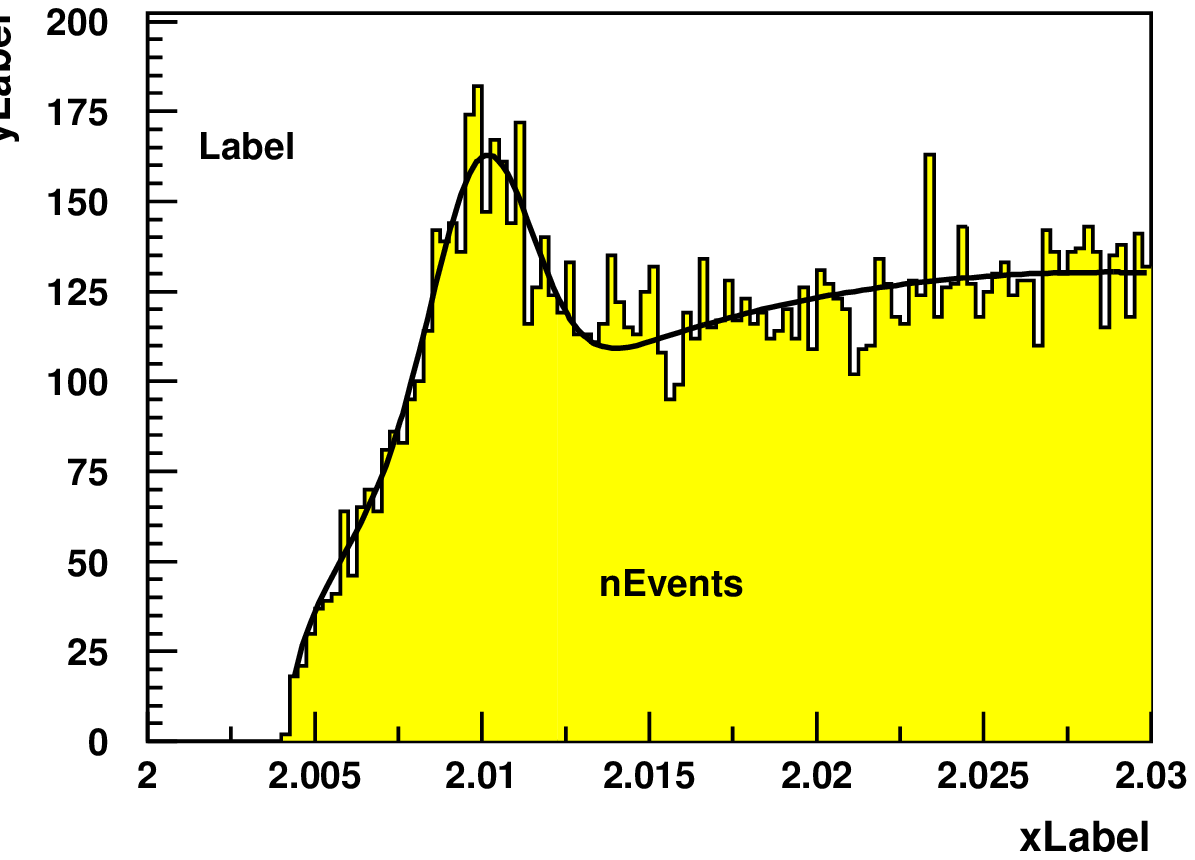} \\
c) $(\KS \pi) \pi$ &
d) $(\KL \pi) \pi$
\end{tabular}
\end{center}
\caption{$D^\ast$ mass plots for the specified $D^0$ decay modes}
\label{fig:signal}
\end{figure}

The resulting $D^\ast$ mass plots for the four $D^0$ decay modes
under study are shown in Figure~\ref{fig:signal}.  For the modes including
$\KS$ these are mass difference plots in the traditional
reconstruction with a very loose cut on the $D^0$ mass (within
$100~\MeV$ from the nominal value).  The distributions for the $\KS$ modes
have been fit to a sum of two gaussians representing the signal and
its tails and a first order polynomial multiplied by a square root
threshold factor to describe the background, while for the signal in
the $\KL$ modes a single gaussian was used.  The central value of all
the gaussians was fixed at the nominal charged $D^\ast$ mass; 
all other parameters were allowed to float.

The yields obtained from the fit are
\begin{eqnarray*}
N(\KS \pi^0) & = & 4715 \pm 91, \\
N(\KL \pi^0) & = & 1839 \pm 101, \\
N((\KS \pi) \pi) & = & 2524 \pm 77, \\
N((\KL \pi) \pi) & = & 1119 \pm 84.
\end{eqnarray*}

They are used to calculate the ratio of branching fractions of $D^0
\rightarrow \KL \pi^0$ and $D^0 \rightarrow \KS \pi^0$
\begin{displaymath}
\frac{\mathcal{B}(D^0 \rightarrow \KL \pi^0)}{\mathcal{B}(D^0
\rightarrow \KS \pi^0)} = \frac{N(\KL \pi^0) / N(\KS \pi^0)}{N((\KL
\pi) \pi)/ N((\KS \pi) \pi)} = 0.88 \pm 0.09 \mathrm{(stat)}
\end{displaymath}


\section{Evaluation of systematic errors}

Due to the similarities in the behavior of the signal $K^0 \pi^0$ and
control $(K^0 \pi) \pi$ modes most of the systematic errors cancel
out.  However some steps in the analysis are different and thus may
have different systematic effect on the result.

We consider the following potential sources of systematic errors which
may not cancel:
\begin{itemize}
\item the effect of the residual difference in $\KL$ lab momentum
spectra, which can lead to a bias in the result if the $\KL$
reconstruction efficiency strongly depends on momentum;
\item systematic errors due to the imperfect fitting function.
\end{itemize}

In order to estimate the bias induced by the residual difference in
$\KL$ lab momentum spectra, an artificial momentum--dependent
``efficiency'' was introduced in $\KS \pi^0$ and $(\KS \pi) \pi$ modes
by applying a weight depending on the $\KS$ lab momentum.  The weight
changed linearly from 0 at $0.6~\GeV/c$ to 1 at $2.5~\GeV/c$,
representing the worst possible case.  The resulting change in the
ratio of yields was less than 3\%.

Systematic errors arising from imperfections of the fitting model
were evaluated by varying the functions describing the signal and the
background: in particular we employed a double gaussian to describe
the signal in $\KL$ modes as well as in $\KS$, and a second order
polynomial multiplied by the square root threshold factor to fit the
background.  The yield ratio remained stable within 6\% under all
variations of the fitting model that were studied.

In addition, the stability of the fit against variation of the
background level was studied by simultaneously varying the cut on
$\cos{\thDK}$ in all the four modes: we estimate the error on the yield ratio
due to this source to be 8\%.

Adding these sources of error in quadrature, we estimate the total systematic
error to be 10\%. Since this error is dominated by the difficulty of
parametrizing the background, there is potential for improvement if the shape
of the background can be simplified. One technique currently being studied
is the replacement of the $M(D^0)$ constraint with a constraint on 
$M^2(D^\ast) - M^2(D^0)$ when forming the \KL\ momentum:
the signal yield is then extracted from the resulting $M(D^0)$ distribution.
As there is no natural threshold in this case, we expect the yield to be
relatively stable under variations of cuts, fit functions and so on.


\section{Conclusion}

In summary, we have demonstrated the viability of this method of
measuring the ratio of $D^0$ decay rates into $\KL \pi^0$ and $\KS
\pi^0$.  Our preliminary measurement of this ratio is
\begin{displaymath}
\frac{\mathcal{B}(D^0 \rightarrow \KL \pi^0)}{\mathcal{B}(D^0
\rightarrow \KS \pi^0)} = 0.88 \pm 0.09 \mathrm{(stat)} \pm 0.09
\mathrm{(syst)};
\end{displaymath}
expressing the result in terms of the rate asymmetry defined
in ~\cite{Golowich}, we find
\begin{displaymath}
\mathcal{A} \equiv \frac{\Gamma(D^0 \rightarrow \KS \pi^0) -
\Gamma(D^0 \rightarrow \KL \pi^0)}{\Gamma(D^0 \rightarrow \KS \pi^0) +
\Gamma(D^0 \rightarrow \KL \pi^0)} = 0.06 \pm 0.05 \mathrm{(stat)} \pm
0.05 \mathrm{(syst)},
\end{displaymath}
which is consistent with unity. At the current level of precision
we are therefore not able to place any strong constraint on the parameters of
$D \rightarrow K \pi$ decays, and so the strong phase $\delta_{K \pi}$. 
However the statistical error will soon improve as more data is accumulated by
the Belle detector; changes of technique to reduce the systematic error
are also being actively studied.


\section*{Acknowledgments}

We wish to thank the KEKB accelerator group for the excellent
operation of the KEKB accelerator.  We acknowledge support from the
Ministry of Education, Culture, Sports, Science and Technology of
Japan and the Japan Society for the Promotion of Science; the
Australian Research Council and the Australian Department of Industry,
Science and Resources; the Department of Science and Technology of
India; the BK21 program of the Ministry of Education of Korea and the
Center for High Energy Physics sponsored by the KOSEF; the Polish
State Committee for Scientific Research under contract No.2P03B 17017;
the Ministry of Science and Technology of Russian Federation; the
National Science Council and the Ministry of Education of Taiwan;
and the U.S. Department of Energy.


\end{document}